\documentclass[twoside]{article}
\usepackage{PRIMEarxiv}

\usepackage[utf8]{inputenc} 
\usepackage[T1]{fontenc}    
\usepackage{hyperref}       
\usepackage{url}            
\usepackage{booktabs}       
\usepackage{amsfonts}       
\usepackage{nicefrac}       
\usepackage{microtype}      
\usepackage{lipsum}
\usepackage{fancyhdr}       
\usepackage{graphicx}       
\graphicspath{{media/}}     

\usepackage[numbers]{natbib}

\usepackage{graphicx} 
\usepackage{import}
\usepackage[table]{xcolor}
\usepackage{soul}
\usepackage{placeins}
\usepackage{float}
\usepackage{adjustbox}
\usepackage{multirow}
\usepackage{caption}
\usepackage{footnote}
\usepackage{threeparttable}
\floatstyle{plain}
\usepackage{listings}
\newfloat{listing}{htbp}{lop}

\usepackage[scaled=0.85]{beramono}
\usepackage[framemethod=TikZ]{mdframed}  
\usepackage{booktabs}   
\usepackage{amssymb}    
\usepackage{pifont}     
\usepackage{afterpage}
\usepackage{todonotes}
\usepackage{amssymb}

\usepackage{pifont}     
\usepackage{afterpage}

\newcommand{\yes}{\textcolor{teal}{\ding{51}}} 

\usepackage{tikz}
\usepackage{hyperref}
\usepackage{xcolor}
\usepackage{fontawesome5}

\definecolor{linkblue}{HTML}{5B6BB5}

\hypersetup{
    colorlinks=true,
    urlcolor=linkblue,
    linkcolor=linkblue,
    citecolor=linkblue
}

\newcommand{\circled}[1]{%
  \begin{tikzpicture}[baseline=(char.base)]
    \node[
      shape=circle, 
      draw=gray,       
      fill=gray,       
      text=white,      
      inner sep=0.5pt, 
      font=\small\bfseries 
    ] (char) {#1};
  \end{tikzpicture}%
}

\newcommand{\projectName}{\texttt{Social-Annotate}}

\usepackage{xcolor}

\title{Social-Annotate: Self-Healing Browser Extension to Annotate and Collect Social Media Data
}

\author{
  Ali Najafi \\
  Sabanci University \\
  Türkiye\\
  \texttt{ali.najafi@sabanciuniv.edu} \\
   \And 
  Ismail Uluturk\\
  University of South Florida \\
  USA \\
  uluturki@gmail.com \\
   \And
  Onur Varol\\
  Sabanci University \\
  Türkiye\\
  \texttt{onur.varol@sabanciuniv.edu} \\
}

\begin{document}
\maketitle

\vspace{-3.5em}
\begin{center}
    \textcolor{linkblue}{\faGlobe}~\href{https://varollab.com/social-annotate}{\textbf{\textcolor{linkblue}{varollab.com/social-annotate}}}
    \qquad
    \textcolor{linkblue}{\faGithub}~\href{https://github.com/ViralLab/social-annotate}{\textbf{\textcolor{linkblue}{github.com/ViralLab/social-annotate}}}
\end{center}
\vspace{2em}

\begin{abstract}
Human-annotated data remains foundational for machine learning and social media analysis. However, traditional data collection often relies on cumbersome pipelines that isolate content from its original source, compromising ecological validity. To address these challenges, we present Social-Annotate, a flexible browser extension that facilitates direct data collection on online platforms. By injecting customizable forms into webpages, the tool captures annotations while users interact with the native environment. \projectName{} offers no-code design interface for the survey forms for non-technical users. Since injecting custom elements directly into host platforms creates a brittle dependency on evolving interfaces, we integrate a self-healing agent powered by large language models. This automated pipeline autonomously detects structural changes, regenerates valid target selectors, and validates them within a live browser environment. Our extensible platform readily supports 12 platforms including social media like $\mathbb{X}$, Instagram, TikTok and P2P messaging platforms WhatsApp and Telegram. \projectName{} significantly reduces data collection overhead and developer maintenance, enabling researchers of all technical backgrounds to focus on data analysis rather than engineering. Moreover, \projectName{} provides an ecosystem for conducting intervention studies by dynamic content manipulation.

\end{abstract}


\maketitle

\section{Introduction}\label{introduction}
Machine learning applications rely heavily on high-quality annotated data to build supervised models and establish reliable ground truth. This is especially true in the domain of social media analysis, where models trained on user behavior and textual content have been historically deployed to detect automated activities and to study online conversations \citep{mitra2015credbank,potthast2018stylometric,shu2020fakenewsnet, varol2017online,yang2019arming}. In recent years, the scope of these models has expanded to address complex information disorders, as malicious actors increasingly leverage autonomous, algorithm-guided social bots to manipulate public opinion and amplify misinformation \citep{arceneaux2025social, ng2025global,malkamaki2026beyond}. Because these automated accounts closely mimic human interactions to steer social discourse, distinguishing them from genuine users requires robust, context-aware classification systems. Consequently, creating rich, accurately labeled datasets has become even more critical for training models to identify algorithmic amplification and preserve the integrity of digital ecosystems \citep{wan2025howdo}.

Similarly, at the post level, developing supervised models for complex content moderation tasks, such as hate speech detection and counter-speech generation \cite{piot2024metahate}, coordinated activity and bot detection \cite{graham2024coordination,zouzou2024unsupervised,varol2017online} requires datasets rich in contextual and linguistic nuance. Identifying toxic or hateful content is notoriously challenging because it often relies on subtle cultural markers, implicit hostility, and evolving in-group or out-group dynamics that simplistic keyword filters frequently miss \citep{persona2025hate, hatebrxplain2025}. To build fair and explainable detection on systems, researchers increasingly depend on human annotators to provide granular rationales and capture diverse community perspectives \citep{hatebrxplain2025}. Furthermore, proactive moderation strategies, such as automated counter-speech generation, depend heavily on high-quality paired datasets of hateful posts and effective, expert-reviewed rebuttals \citep{damo2025beating, enhancing2025counter}. Without tools that facilitate the rigorous, in-context annotation of these subjective interactions, models remain susceptible to representational biases and struggle to generate culturally sensitive interventions.

To build annotated datasets for model training, researchers are often faced with inefficient tasks of inspection and labeling. In most data collection studies, content to be inspected is often shared as a collection of URLs contained in platforms such as online forms and spreadsheets. This content is then inspected manually by following the URL directly on a web browser, or the content is isolated and extracted from its source on a separate data collection platform. Crowd-sourcing approaches can help scale the annotation task; however, the cost of high-quality annotated data is still expensive and demanding since the performance of non-expert annotators requires further validation or larger samples \cite{hsueh2009data}. 
Tools like AnnotHate designed to speed up these processes and demonstrated the efficiency in terms of time spend on annotation \cite{jikeli2024annotating}.
Furthermore, the specific design of the crowd-sourcing task directly influences the interpretations and labels provided by untrained workers, meaning these varying interpretations must be carefully gathered and validated \cite{yung2025crowdsourcing}. 


Another significant risk researchers face is the content removal or changes in the platform regulations.
In this post-API era, stable and official access points are disappearing \cite{freelon2018computational}. 
While major platforms historically offered documented APIs for academic research, the ``APIcalypse'' has seen companies like $\mathbb{X}$/Twitter and Facebook restrict or monetize access, effectively privatizing public data \cite{bruns2021after}. 
This shift poses a severe risk to scientific equity, as high-quality data becomes a luxury available only to well-funded institutions, while others are left to navigate a fragmented landscape of ``fragile'' data \cite{davidson2023platform}. Similarly, Assenmacher et al. argues datasets available for misinformation researchers suffer significant losses due at the rehydration steps due to content deletions \cite{assenmacher2023end}.

While the data collection approaches described so far appear feasible, they often require a constant back-and-forth between the browser and forms, or they display data that have been previously extracted alongside forms on a separate platform, isolated from the original online source of the content. Both of these approaches introduce significant overhead in the collection, preparation, and access of content to be annotated. Furthermore, these approaches isolate the content annotation task from the natural use of the online platform, which may impact user experience and introduce or remove biases compared to regular users. Recommendation algorithms, the sequence content is loaded, and content layout on the page are some examples of these sources of influence inherent to each platform.

This disconnect highlights a lack of ecological validity in standard annotation pipelines. In human-computer interaction, ecological validity refers to the extent to which a testing environment accurately reflects the real-world conditions where a tool is naturally used \citep{brunswik1956perception}. Traditional annotation tools often require users to shift away from their standard workflows into isolated applications, which can introduce artificial cognitive load and skew data collection \citep{carter2008exiting}. Recent studies underscore that embedding data collection within the natural environment of social media users significantly enhances this validity \citep{aridor2024experiments}. By designing our annotator as a browser extension integrated directly into the target platform, our approach preserves ecological validity. Users engage with the tool natively in the original environment. This ensures that the captured interaction patterns and annotations represent real-world application, accounting for the native layout and algorithmic behaviors that influence user perception.

Furthermore, while certain aspects of the annotation task can be automated or augmented to reduce overhead, these automated approaches often require implementing custom data pipelines or tools. This introduces a cost of entry that excludes researchers with limited technical know-how or financial resources and makes studies of smaller scope less feasible in terms of effort or financial cost.  

Integration of data collection tools directly into browser extension preserves ecological validity; however, it introduces significant technical challenges and fragility for the tool. Because injecting custom elements creates a brittle dependency on the proprietary Document Object Model (DOM) structures of host platforms, even minor interface updates can break the underlying annotation tool. To address this maintenance challenge, we introduce a self-healing agent powered by large language models (LLMs). By utilizing LLM APIs to analyze structural changes, the agent autonomously repairs broken target selectors to align with the updated interface. This capability ensures the system remains robust despite the continuous evolution of online platforms, preserving reliability and reducing the need for constant developer maintenance.

Here, we present \projectName{}, an extendable and configurable browser extension for data collection, conducting intervention studies and data annotation, to achieve following goals:

\begin{itemize}
    \item Bringing online data sources and data collection platform together to speed-up the annotation process.

    \item Survey forms that are easy to customize for non-technical users, and extendable for other websites.

    \item Datasets collected for a study can be exported to JSONL files, or submitted to an API endpoint automatically. If a platform contains media objects such as image and video, they can also be downloaded for further analysis.

    \item Browser extension can inject survey question directly into websites so that annotation can be collected while interacting with the source platform naturally, or manipulate the displayed content for controlled studies. 

    
    \item Autonomous adaptation to UI changes via self-healing LLM agents to ensure long-term stability and reduce developer maintenance.
\end{itemize}

\section{System Design}
\begin{figure}[!t]
    \centering
    \includegraphics[width=\linewidth]{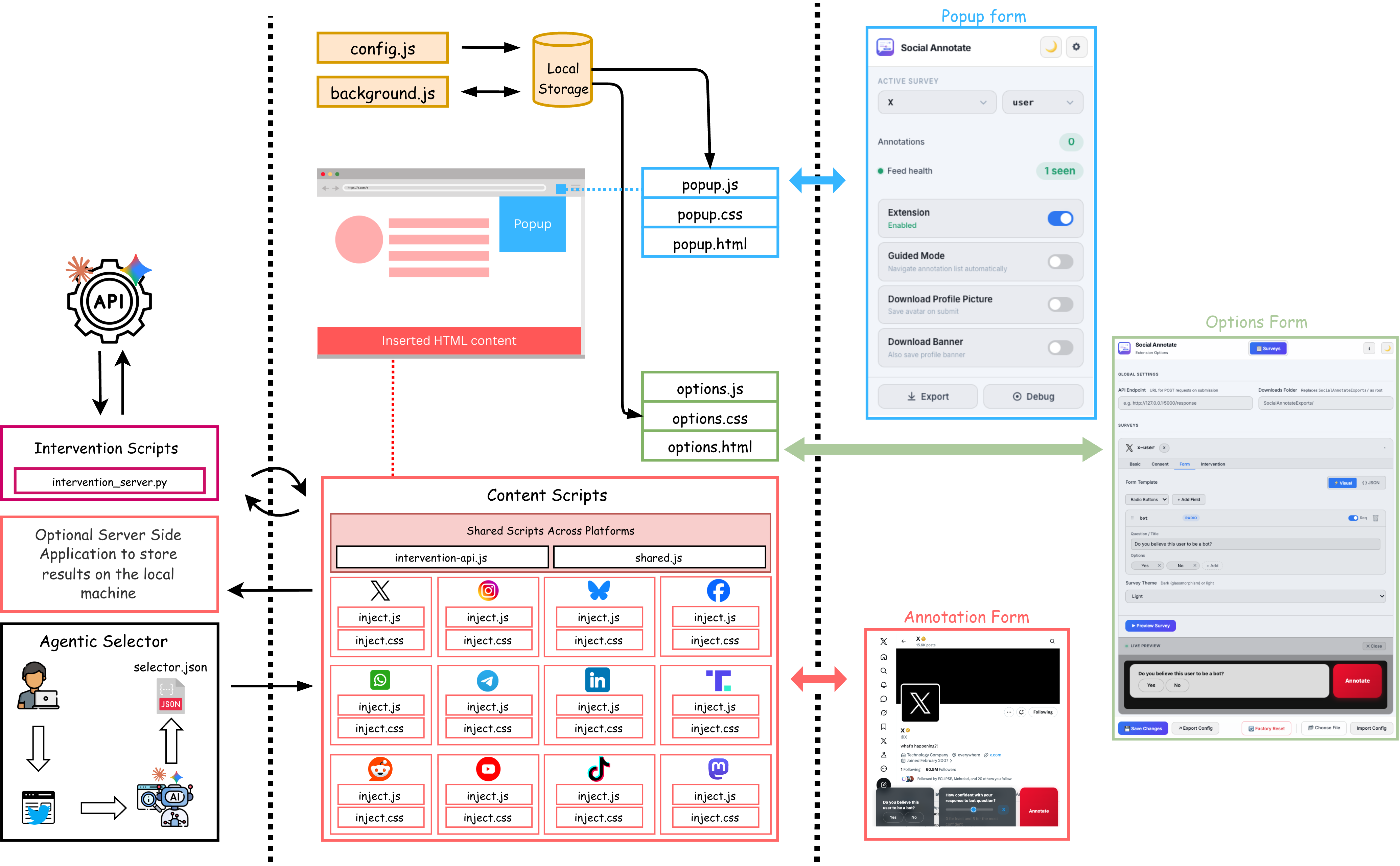}
    \caption{\textbf{\projectName{} platform design schematic.} Components of the extension presented in the middle panel and their UI designs shown on the right panel.}
    \label{fig:schema}
\end{figure}

We build our browser extension following the design guidelines and best practices provided
by Google Chrome Extensions.\footnote{\href{https://developer.chrome.com/docs/extensions}{https://developer.chrome.com/docs/extensions}}
This makes our extension compatible with Chromium-based browsers like Brave, Vivaldi for privacy.
Figure~\ref{fig:schema} presents the high-level representation of the extension. Elements of the extension are presented in the middle panel. We implement a separate content script for each social media platform, and only inject the relevant scripts to the page. Content scripts inject the surveys to the page, where the injection location and the survey itself can be configured. Users can interact with the extension using ``popup'' and configure it using ``options'' pages.
Extensions architecture mainly consists of following units:

\begin{itemize}
    \item \textbf{manifest.json}: Provides information about the extension, functional elements of it,and necessary permissions. Required for every extension by Google Chrome. This configurations allows us to request permission to edit certain websites by injecting style and scripts.

    \item \textbf{Background scripts}: Handle background operations and listens for events interfacing with the extension.

    \item \textbf{Content scripts}: Main functional elements of the extension. Facilitates the interface between the page content, extension, and browser APIs.

    \item \textbf{Popup page}: A simple interface to control the extension, such as enable/disable, annotation statistics, inspecting the selectors usability and switching between different collection tasks.

    \item \textbf{Options page}: A page for configuring and customizing the extension to change the behavior of certain aspects. Configurations on this page can be exported to or imported from a file.

\end{itemize}

Moreover we provide two additional components that enriches the extension usability:

\begin{itemize}
    \item \textbf{Self-Healing Agent}: Provides a functionality to update extension for continuous support in case the UI changes on the target platforms. We offer more detailed description of this feature on Section ~\ref{self_healing_agent}.
    \item \textbf{Content Manipulation Agent}: Provides an interface to make API calls to external APIs or LLM models such as Gemini and Claude so that the study organizers can make an intervention on the displayed content before starting the annotation process.
\end{itemize}

An important feature of our extension is its ``no-code'' aspect to conduct studies and changing questions for the study. Users can easily change the survey format.
We aim to be accessible for all users, especially those with limited programming experience. \projectName{} already supports 12 platforms listed on Table \ref{tbl:platforms} and example screenshots can be seen in the Supplementary Information (see Fig.\ref{fig:survey_examples}). Surveys can be customized in the options page as described in detail on Section \ref{sec:annotation-forms}, and entire configurations including customized surveys can be exported as a configuration file that can be shared with study participants for easy onboarding.

Advanced users can also implement their own content scripts for extending to different platforms. Using this additional functionality, users can design their own data collection strategy and manipulate content for controlled studies. We are providing recipes on how to easily extend \projectName{} for supporting other platforms in our Github repository.

For conducting studies that require intervention by manipulating the displayed contents, the organizers can provide either mappings from original content to the manipulated content or use external REST API using the \texttt{Content Manipulation Agent} which does API calls and has customizable prompting options. In Section \ref{intervention}, we present this feature in more details.

\section{Annotation Forms}
\label{sec:annotation-forms}

Surveys can contain an arbitrary number of questions of various types, tailored to the specific needs of a study. We utilize JSON schemas as templates to facilitate the easy configuration and sharing of custom survey forms. This approach provides access to a comprehensive range of form elements and controls without requiring any HTML or Javascript coding, empowering researchers with limited technical expertise to create custom forms easily, supported by the jsonform package \cite{jsonform}. 
When a user navigates to a URL matching a target domain, the corresponding survey is dynamically injected into the webpage. The extension supports platform-specific surveys, which also dictate the specific HTML elements to which the forms are attached. The system supports two primary survey types: account surveys (to annotate users) and post surveys (to annotate messages). For example, on a platform like $\mathbb{X}$ or Bluesky, researchers can deploy distinct forms for annotating individual users or specific tweets on a profile or timeline.

\begin{figure}[t!]
    \centering
    \includegraphics[width=\linewidth]{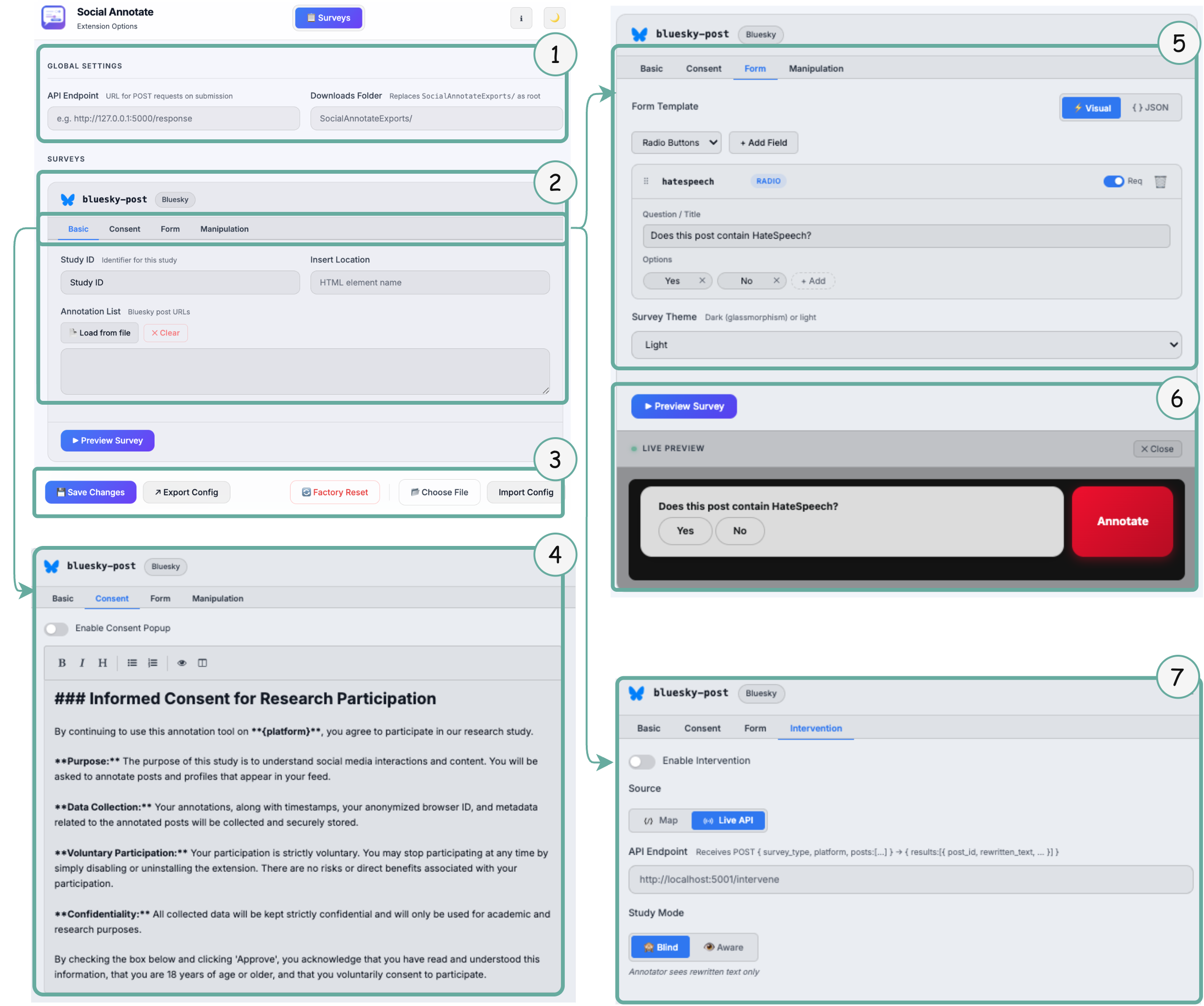}
    \caption{\textbf{\projectName{} options page.} The figure presents different forms or interfaces to setup a study and configure experiment. (1) API endpoint configuration for POST requests, and media download path configuration, (2) Basic configuration panel consist of \texttt{Study ID}, annotation list management with file upload capability and HTML element insertion location setting, (3) configuration management controls including save changes export, factory reset, and file operations, (4) informed consent popup with customizable form text, (5) form template selection with visual/JSON export options, (6) preview of survey form, (7) Content Manipulation Panel with Blind/Aware Study mode setting along with live API endpoint configuration and file upload capability of the targeted content.}
    \label{fig:options_page}
\end{figure}

Figure~\ref{fig:options_page} highlight different configurations and features of the options page for users to design their study and collect results. The figure also highlight parts from \circled{1} to \circled{7} of the options page for different settings and configuration forms.

In part \circled{1} of the options page, users can configure data storage and export settings. The application captures account and post data alongside the survey answers, temporarily saving them in the browser's local storage. Users can change the default download directory for \projectName{} to export this data at any time. Alternatively, users can set an API endpoint if they are using a backend server to capture live annotation results via direct HTTP requests. This option is ideal when multiple annotator working on different machines. To unify responses across platforms and ease integration with various downstream tools, we adopted the Account and Post format of the Social Media Data Toolkit (SMDT) \cite{najafi2026social}.

In part \circled{2}, there are four tabs display different setting subsections. In the \textit{Basics} subsection, users can: i) set a study ID, ii) add a targeted annotation list via editing or uploading a file, and iii) manually change the insert location. 
Researchers can then configure their surveys by adding or removing questions in various formats, including radio buttons, checkboxes, ranges, and text inputs on tab called \textit{Form}. This interactive form displayed in form \circled{5}. Once the survey is ready, users can click the Preview Survey button in the part \circled{6} to observe how the form will appear.

As shown in the part \circled{5} of the Figure~\ref{fig:options_page}, users have multiple ways to configure these forms: i) they can design questions interactively using the application's options panel, ii) directly input the JSON schema code, or iii) import a pre-configured file prepared by the annotation organizer. Part \circled{6} presents an preview of a survey form designed to support hate-speech detection research on Bluesky. 

Based on study needs, there may be a requirement to manipulate the displayed content. For instance, rewriting with a more polite language or censoring certain information in the content. In part \circled{7}, researchers can enable content manipulation in two different settings: i) Blind: where the user will not know whether the content was manipulated, or ii) Aware: where the annotator will see a button with a text \textit{Show Original} to view the original content.

Finally, to ensure ethical compliance in human-subjects research, \projectName{} includes a configurable consent form on a tab called \textit{Consent}. Recent scholarship emphasizes that establishing explicit informed consent is an essential requirement to protect crowdworkers and resolve the ethical ambiguity surrounding their status as human subjects \cite{hawkins2023ethical,kaushik2022resolving}. 
Because ethical guidelines and Institutional Review Board (IRB) requirements vary significantly across institutions, a static disclosure is insufficient. By enabling and editing the consent subsection, researchers can tailor the agreement to their exact study parameters. For this reason, in the section \circled{4}, study organizers can prepare a consent form using the provided markdown editor.  When annotators visit the target platform for the first time, they will be required to review and explicitly approve this form prior to starting their tasks. 

After any amendments, users can save their changes, discard them using the factory reset button, or import their configuration file as shown in \circled{3}.

\section{Self-Healing Agent}\label{self_healing_agent}

Social media platforms continuously update their front-end codebases, altering the Document Object Model (DOM) structures that browser-based annotation tools depend on to locate and inject components. 
When a selector becomes stale, tools silently fail to inject survey forms, producing data gaps that are difficult to detect retroactively. Manual correction requires developers to inspect live HTML for each affected platform. This is a process that does not scale across rapidly evolving sites. To address this challenge, we designed a Self-Healing Agent. 
This automated pipeline combines large language model (LLM) reasoning over raw HTML with end-to-end browser-level validation to autonomously regenerate a platform's selector configuration whenever its DOM structure diverges from a known-good specification.

The agent's architecture consists of three core components. First, a platform abstraction layer encapsulates the site-specific knowledge required for each supported network (e.g., $\mathbb{X}$ / Twitter, Instagram, Bluesky, Telegram), which is a typed selector schema, a validation function, and a natural-language prompt template. 
Schemas are defined using Pydantic, allowing field-level descriptions to act as structured constraints for the LLM. Second, an LLM client handles structured selector extraction, interchangeably supporting backends like Claude \cite{anthropic2024claude3} and Gemini \cite{gemini2023family}. These models are invoked through constrained output mechanisms, such as tool use or structured response schemas, that guarantee the output conforms to the expected schema without post-hoc parsing. Third, a live browser environment built on Playwright loads the proposed selectors into a Chromium instance running the unpacked extension. This allows the agent to observe real injection behavior rather than simulating it. Throughout this process, offline HTML analysis is used for initial structural pre-checks and for validating candidate selectors before the heavier browser environment is launched.

\begin{figure}[!t]
    \centering
    \includegraphics[width=\linewidth]{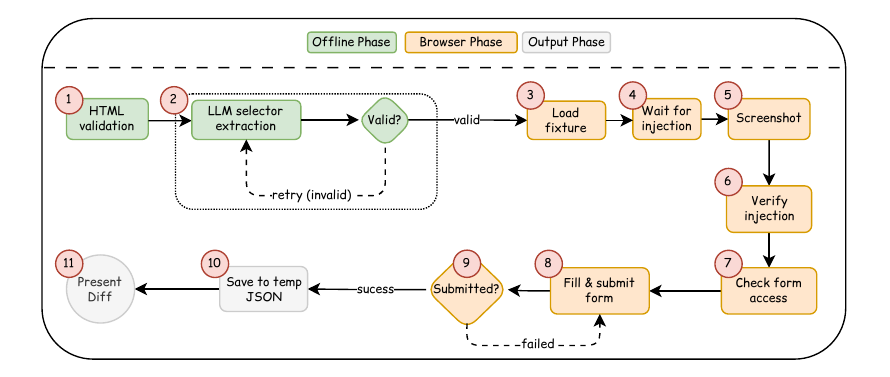}
    \caption{\textbf{Architecture of the self-healing agent pipeline for dynamic survey injection.} The workflow is divided into four main stages: Validation, Extraction, Browser, and Output. Initially, target DOM elements undergo HTML validation before the LLM extracts the necessary selectors. The extraction process includes a retry mechanism to ensure selector validity. In the Browser environment, the fixture is loaded, and the system uses screenshots to verify the form injection. The agent then accesses, fills, and submits the annotation form, recursively retrying upon failure. Upon success, the data is saved to a temporary JSON file and evaluated through a presentation diff.}
    \label{fig:self_healing_agent}
\end{figure}

As illustrated in Figure~\ref{fig:self_healing_agent}, the self-healing pipeline operates in three main phases:

\begin{enumerate}
\item \textbf{Offline Phase:} The agent first parses the HTML fixture to estimate the detectability of post elements using known patterns, flagging instances where the DOM has changed substantially (Step \circled{1}). The HTML is then pruned to reduce token volume and submitted to the LLM alongside the selector schema and context. The LLM returns a candidate selector set, which is immediately validated offline against the fixture (Step \circled{2}). If validation fails, the error is fed back into the prompt, allowing the model to self-correct without human intervention.

\item \textbf{Browser Phase:} Proposed selectors are written to the extension's runtime storage, and the fixture is opened in a live Chromium instance (Step \circled{3}). The agent triggers the content script's mutation observer by scrolling the viewport, scanning for post elements as they render (Step \circled{4}). After capturing a visual record of the browser state (Step \circled{5}), the agent verifies the successful injection of survey containers and sandboxed iframes (Step \circled{6}), ensuring the internal forms are accessible and renderable (Step \circled{7}). It then simulates user interaction by submitting a test survey (Steps \circled{8} and \circled{9}). Critically, this phase goes beyond confirming form submission; it inspects the stored annotation record to verify that the extracted selectors accurately capture post identifiers, timestamps, and engagement metrics.

\item \textbf{Output Phase:} The validated selector configuration is staged for deployment rather than written directly to production (Step \circled{10}). The system generates a field-level diff against the current production configuration, providing operators with a focused summary of the changes (Step \circled{11}). Final deployment requires explicit operator approval, maintaining human oversight as a mandatory safety gate.

\end{enumerate}

With this Self-Healing Agent, users can supply their HTML and provide supplementary context to guide the agent's behavior. This mechanism allows advanced users to inject specific domain knowledge prior to execution. For example, a user can explicitly instruct the agent that a target platform, such as WhatsApp, does not support reposting actions, ensuring the resulting selectors align with the platform's actual capabilities.

Since we cannot anticipate the future changes of current platforms, we develop ad-hoc scenarios to test capabilities and blind-spots of the self-healing agent.
To evaluate the effectiveness of the self-healing agent on a given webpage, we quantify how frequently the agent succeed the task of extracting the selectors needed for survey-form injection and data collection. 
For $\mathbb{X}$, we collected historical pages from the Internet Archive's Wayback Machine. The agent achieved full annotation functionality across all five evaluated $\mathbb{X}$/Twitter snapshots (2010, 2014, 2017, 2020, and 2026), with complete metadata capture in four of the five cases. The only issue encountered on the 2010 snapshot, where the absence of standardized timestamp attributes prevented \texttt{created\_at} extraction, despite successful survey injection and submission. Bluesky was similarly evaluated using Wayback Machine snapshots from 2023 and 2026, both of which passed successfully.

For the remaining platforms, archival snapshots were not available. Instead, we supplied a saved snapshot of each platform's current page without exposing the agent to any predefined selectors. Under these conditions, the agent successfully extracted selectors and injected surveys for Reddit (feed and comment thread), Mastodon, WhatsApp, Telegram, and Truth Social. LinkedIn was excluded from evaluation due to an LLM input-size limitation: the platform's saved page HTML exceeds the practical generation window of the model, causing extraction to time out. We did not evaluate the agent on platforms such as Instagram, where DOM elements inserted though JavaScript as part of the page life-cycle. For example, the comment section appears only after the comment button is clicked, although such click actions could also be automated on a live system, we cannot test this from a static HTML snapshots.

\section{Intervention Studies}
\label{intervention}

For more sophisticated use cases and advanced users, we implemented a module to conduct intervention studies. In this module, \projectName{} can send user or post data to a REST API and collect responses to update corresponding fields on the website.

Since the possibilities for conducting intervention studies are endless, we offer a boilerplate REST API that replaces posts with a generic \textit{lorem ipsum} text and updates user profile metrics with a numerical value of $-1$ (see examples in Fig.\ref{fig:intervention_examples}). 
Users can extend this method to change specific entities of the posts like hashtags, URLs, and mentions or the entire text can be rewritten by an NLP model. This feature is ideal for studying how language use can affect user behavior or how profile metrics influence engagement. 

An alternative approach that does not require a REST API is also available. This approach relies on a precomputed mapping between original and manipulated content. This mapping file can be uploaded, and the changes are rendered directly from the file while the page is loading.

Since some use cases may require signaling whether or not the content is manipulated, we offer two study modes: Blind and Aware. In \textit{blind mode}, users are not aware that the content is being manipulated, allowing the experiment to collect responses to that manipulated content. When the tool is set to \textit{aware mode}, participants have the control to switch between the original and manipulated content. Lastly, this feature can also be used to scrape all the content user sees by simply running studies in \textit{blind mode} and not changing any of the content.

\section{Conclusion}
\label{conclusion}

In this work, we introduced \projectName{}, an extendable browser extension designed to streamline online data collection and human annotation. 
Traditional annotation pipelines often force researchers into isolated environments, sacrificing ecological validity. By injecting survey elements directly into target platforms, our tool preserves the native user experience and captures authentic interaction patterns. 
This design considerations address limitations due to data access and complexities of conducting annotation studies at scale.

Efforts to collect online data utilized browser extensions on studies monitoring user behavior \cite{capra2010hci, achmann2024preserving}, crowdsource tags \cite{kaur2014scholarometer}, news consumption \cite{chakraborty2016stop}, and social bots \cite{khayat2019vassl}. \projectName{} builds upon these methods to address sampling biases that frequently occur in domains requiring diverse behavioral datasets. Crowdsourcing the annotation task allows real users to label accounts they observe directly in their own social media timelines, surfacing a more representative set of data.
Promising direction to support research on social media incorporate collection of digital trace data through mobile applications, browser extensions, and screen tracking \cite{van2022promises,ohme2024digital}.

Beyond behavioral tracking, browser extensions have proven invaluable for real-time content verification and structured data extraction. For instance, the InVID Verification Plugin provides journalists and researchers with a comprehensive suite to debunk misinformation through in-browser reverse image searches and forensic video analysis \cite{teyssou2017invid}. Similarly, tools like the Instant Data Scraper use heuristics to automatically identify and extract tabular data from web pages \cite{instantdatascraper2026}. However, such tools often struggle to extract structured data from dynamic platforms like $\mathbb{X}$. Ultimately, these applications highlight the broad utility of browser-based interfaces in reducing the operational friction of online research. This aligns closely with the primary objective of \projectName{} to empower researchers regardless of their programming expertise.  

To further support users of the \projectName{}, we address the important challenge of platform changes and robustness of the experimental pipelines.
To address the inherent fragility of DOM-dependent extensions, we developed a self-healing agent powered by large language models. This automated pipeline autonomously detects structural changes, regenerates valid selectors, and validates them in a live browser environment. This approach significantly reduces developer maintenance and ensures long-term stability across rapidly evolving platforms.


There are promising directions remain for extending the capabilities of \projectName{}. Primarily, we intend to expand the extension's existing intervention features to support more sophisticated content manipulation mechanisms. Currently, the tool allows researchers to manipulate displayed content through predefined configurations. However, the modular structure of the tool allow researchers to design and integrate their own models to conduct more complex studies. In future iterations, we plan to integrate local, in-browser artificial intelligence models to significantly broaden these display intervention possibilities.
To achieve this without compromising user privacy or incurring external API costs, one can leverage emerging client-side AI capabilities, such as window.ai\footnote{\href{window.ai}{window.ai}} which is a Chrome's built-in AI, alongside execution frameworks like \texttt{Transformers.js} and \texttt{WebLLM}. Integrating these models will enable complex text and content transformations to be processed entirely on the user's local machine. This on-device AI integration will unlock advanced intervention scenarios for researchers, including: i) Contextual Content Alteration by using local LLMs to intelligently modify, rewrite, or obfuscate specific user descriptions, profile details, or post content based on experimental parameters on browser. No need to mention we are currently supporting interventions for user accounts and posts. ii) Engagement Metric Adjustment by systematically altering or masking engagement statistics, such as reposts, likes, or view counts, to test theories on social proof and algorithmic amplification.

It is worth noting that the self-healing agent described in this work
relies on the assumption that selectors are sufficient to identify the
injection location and capture the intended information. We anticipate
that AI agents will become increasingly capable, enabling them to
generate reliable code that captures the underlying logic of a webpage
and thereby reducing the need for human supervision in preparing the
tool for use. In the meantime, the growing availability of coding
agents already lowers the barrier to adapting the extension to new
platforms and use cases.

By utilizing local AI to manage the alteration of these specific elements, researchers can effectively mitigate social influence bias during the annotation process without relying on external servers. Ultimately, these advanced, privacy-preserving manipulation capabilities will provide a more robust and flexible environment to study human behavior under controlled, yet ecologically valid, online conditions.

\section*{Acknowledgments}

This work is partly supported by Technological Research Council of Türkiye (grant: 222N311) and EU HORIZON project ECLIPSE (grant: 101225823).

\bibliographystyle{unsrt}  
\bibliography{refs}

\clearpage
\section*{Supplementary Information}
\setcounter{figure}{0}
\renewcommand{\thefigure}{S\arabic{figure}}
\setcounter{table}{0}
\renewcommand{\thetable}{S\arabic{table}}

\begin{figure}[h]
    \centering
    \includegraphics[width=\linewidth]{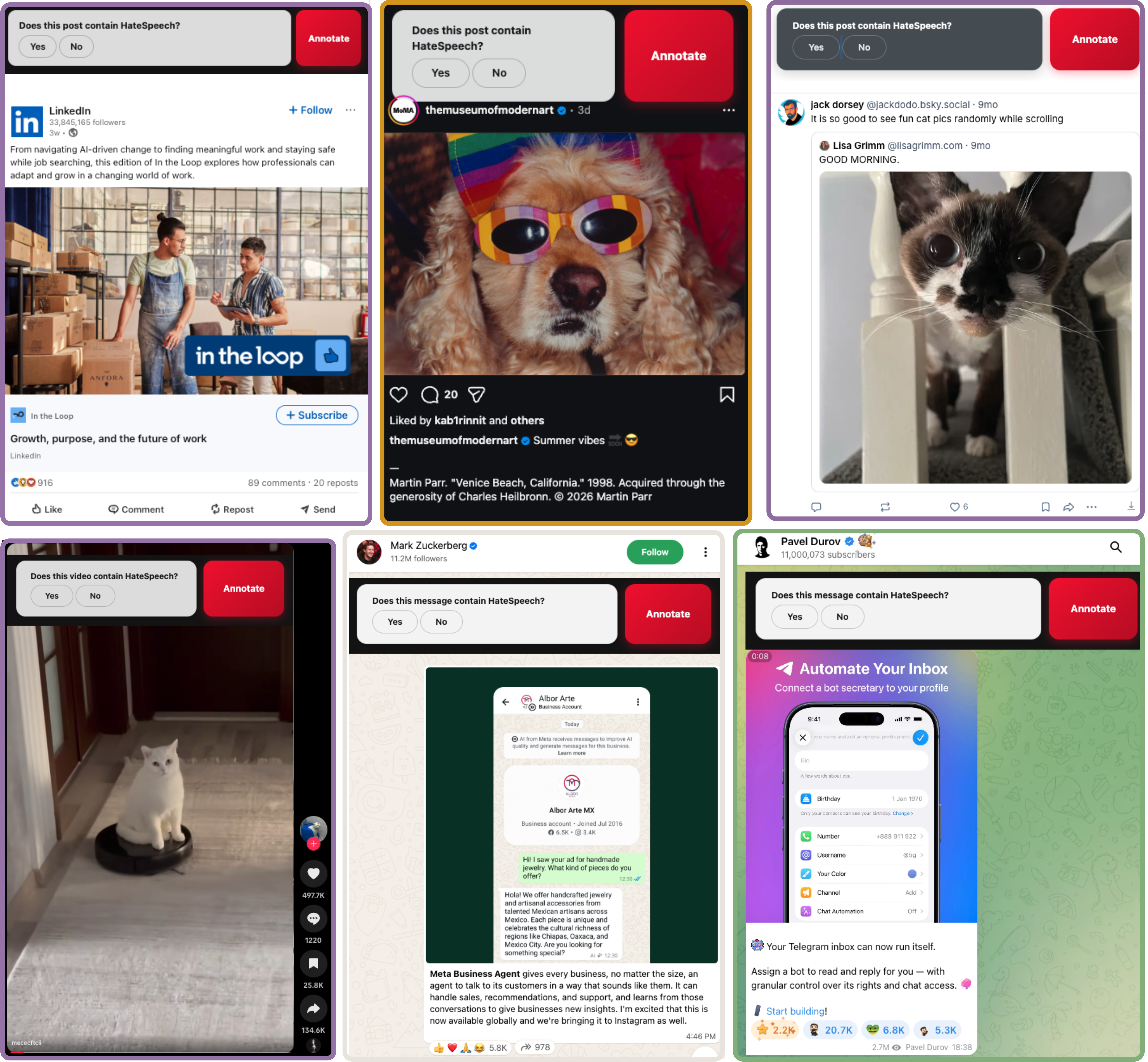}
    \caption{\textbf{Social-Annotate deployed across diverse platforms}. Hate-speech survey form (``Does this post contain HateSpeech?'') is injected natively into six social media environments (LinkedIn, Instagram, and Bluesky in the top row; Telegram, WhatsApp, and $\mathbb{X}$ in the bottom row), demonstrating the extension's cross-platform generality. In each case the annotation form is anchored to platform-specific DOM elements while preserving the original content, layout, and engagement cues, illustrating how data collection is embedded directly within the user's native browsing experience.}
    \label{fig:survey_examples}
\end{figure}

Figure \ref{fig:survey_examples} presents screenshots from different platforms. Currently the extension supports the following platforms and content types summarized on Table \ref{tbl:platforms}. 

\begin{figure}[h]
    \centering
    \includegraphics[width=0.95\linewidth]{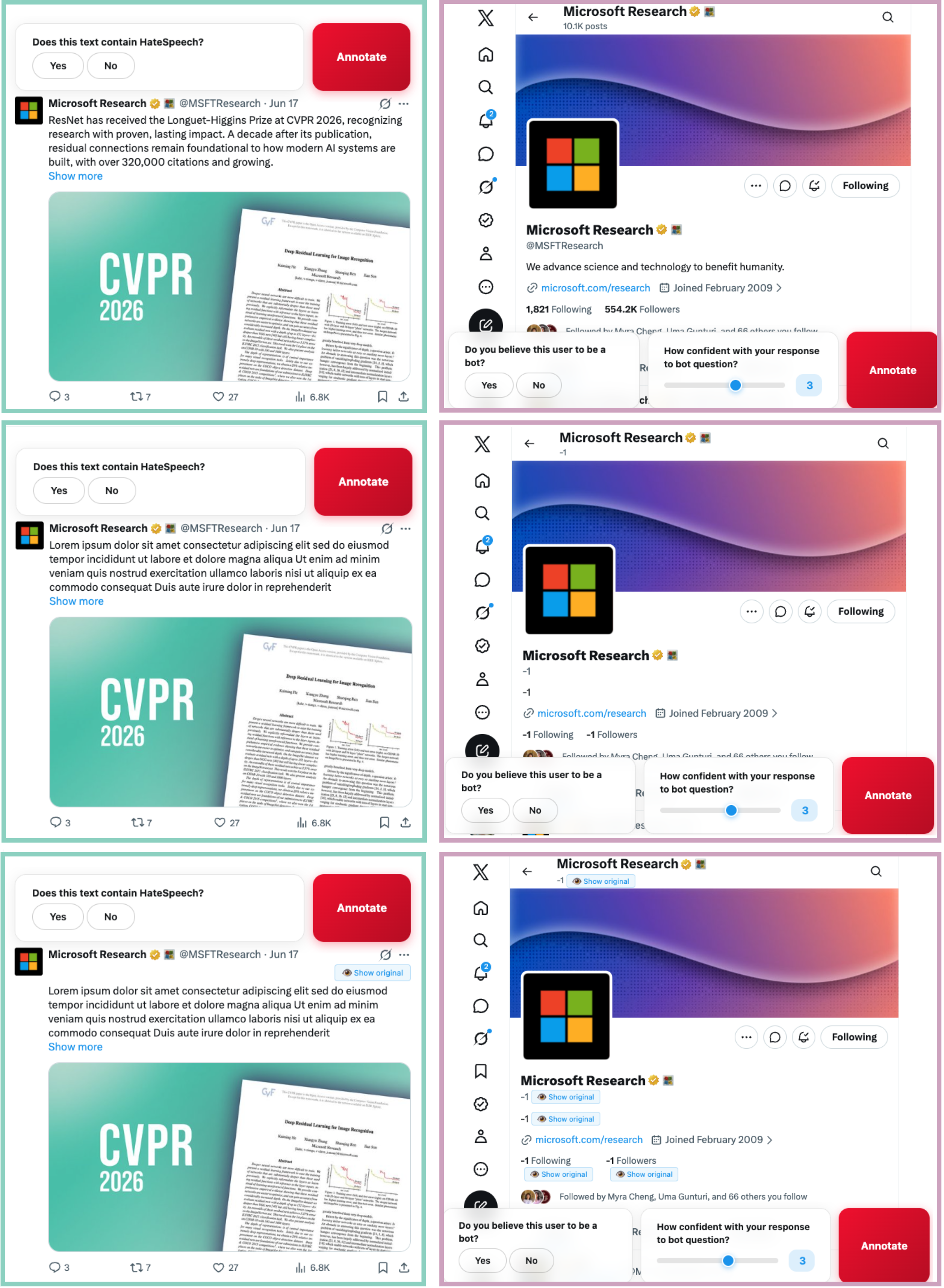}
    \caption{\textbf{Intervention on content and profile examples.} The images present screenshots taken from a sample platform for a post (left) and a user (right). Different modes of intervention tool presented in rows: original content (top), manipulated content in \textit{blind mode} (middle) and \textit{aware mode} (bottom).}
    \label{fig:intervention_examples}
\end{figure}

Figure \ref{fig:intervention_examples} presents screenshots taken for simple intervention studies as an illustration of this module's capabilities. Simply the content manipulation service replaces post text with \textit{lorem ipsum} text and updates numerical profile metadata with $-1$.

\begin{table}[h]
\centering
\caption{\textbf{List of supported platforms.} \projectName{} currently offers functionalities for the listed platforms on Github. Intervention study feature is implemented on some of the platforms for account and post fields.}
\label{tbl:platforms}
\rowcolors{2}{gray!10}{white}
\begin{tabular}{l|ccccc|cc}
    \toprule
    & 
    \multicolumn{5}{c|}{\textbf{Data Collection}} & \multicolumn{2}{c}{\textbf{Intervention}} \\
    \multirow{-2}{*}{\textbf{Platform}} & \textbf{Posts} & \textbf{Accounts} & \textbf{Comments} & \textbf{Videos} & \textbf{Reels} & \textbf{Account} & \textbf{Post}\\
    
    \midrule
    $\mathbb{X}$ / Twitter & \yes & \yes & & & & \yes & \yes\\
    TikTok                 &  & \yes & & \yes & \yes\\
    Instagram              & \yes & \yes & \yes & & \yes & \yes & \yes\\
    Facebook               & \yes & \yes & & & & \yes& \yes\\
    Bluesky                & \yes & \yes & & & & \yes & \yes\\
    Mastodon               & \yes & \yes & & & & \yes & \yes\\
    Truth Social           & \yes & \yes & & & & \yes & \yes\\
    \midrule
    LinkedIn               & \yes & \yes & & & & & \yes\\
    Reddit                 & \yes & \yes & \yes & & & & \yes \\
    YouTube                &  & \yes & \yes & \yes & \\
    \midrule
    WhatsApp               & \yes & & & & & & \yes\\
    Telegram               & \yes & \yes & & & & & \yes\\
    \bottomrule
\end{tabular}
\end{table}

\end{document}